\documentclass[conference,10pt]{IEEEtran}
\usepackage{epsfig,epsf,rotating,setspace,latexsym,amsmath,amssymb,amsfonts,bm,theorem,subfigure,epstopdf,cite,authblk,bbm}
\usepackage{algorithm}
\usepackage[noend]{algpseudocode}
\usepackage{color}
\usepackage{mathtools}
\usepackage{multirow}
\usepackage{soul}

\algrenewcommand\algorithmicforall{\textbf{foreach}}
\algrenewcommand\algorithmicindent{.8em}

\newtheorem{theorem}{Theorem}

\newtheorem{remark}{Remark}

\newtheorem{example}{Example}

\IEEEoverridecommandlockouts
\allowdisplaybreaks

\begin{document}
\title{HetDAPAC: Distributed Attribute-Based Private Access Control with Heterogeneous Attributes}

\author{Shreya Meel \qquad Sennur Ulukus\\
\normalsize Department of Electrical and Computer Engineering\\
\normalsize University of Maryland, College Park, MD 20742 \\
\normalsize {\it smeel@umd.edu} \qquad {\it ulukus@umd.edu}}

\maketitle

\begin{abstract}
Verifying user attributes to provide fine-grained access control to databases is fundamental to an attribute-based authentication system. In such systems, either a single (central) authority verifies all attributes, or multiple independent authorities verify individual attributes distributedly to allow a user to access records stored on the servers. While a \emph{central} setup is more communication cost efficient, it causes privacy breach of \emph{all} user attributes to a central authority. Recently, Jafarpisheh~et~al.~studied an information theoretic formulation of the \emph{distributed} multi-authority setup with $N$ non-colluding authorities, $N$ attributes and $K$ possible values for each attribute, called an $(N,K)$ distributed attribute-based private access control (DAPAC) system, where each server learns only one attribute value that it verifies, and remains oblivious to the remaining $N-1$ attributes. We show that off-loading a subset of attributes to a central server for verification improves the achievable rate from $\frac{1}{2K}$ in Jafarpisheh~et~al.~to $\frac{1}{K+1}$ in this paper, thus \emph{almost doubling the rate} for relatively large $K$, while sacrificing the privacy of a few possibly non-sensitive attributes.
\end{abstract}

\section{Introduction}
Attribute-based encryption \cite{watersABE} is an authentication strategy wherein a central authority verifies the \emph{attributes} of a user, in order to grant access to certain data. Here, a user, identified by its set of attributes, wishes to access its designated record by getting its attributes verified by an authority. If instead of a single trusted authority, the task of verifying attributes is distributed among multiple independent authorities, the resulting variant is known as multi-authority attribute-based encryption \cite{chaseMABE}. Such systems are studied in several works using cryptographic primitives such as bilinear mapping, encryption algorithms and hashing \cite{jungprivateMABE, wangABE, saeedPH3R}. All these works provide only computational privacy guarantees to the user.

Reference \cite{aliDAPAC} is the first paper to formulate an information theoretic framework for the multi-authority attribute-based encryption system in the form of the distributed attribute-based private access control (DAPAC) system. A DAPAC system comprises multiple non-communicating authorities (servers) and a single user, where the user requesting its data record from the authorities, is uniquely identified by its attribute vector. While the security and privacy constraints are similar to those of symmetric private information retrieval (SPIR) \cite{jafarSPIR}, the user has access to \emph{non-replicated} databases in DAPAC, unlike the SPIR setup. PIR \cite{jafarPIR, ulukusPIRLC} from non-replicated databases has been studied in \cite{banawanNONREPIR, ravivNONREPIR, jafar4starPIR} under various graph structures. 

In this work, we explore a hybrid $(N,D,K)$ heterogeneous DAPAC (HetDAPAC) system wherein a subset of $D$ attributes are verified by $D$ distributed authorities (one attribute per authority), while a single central authority verifies the remaining $(N-D)$ attributes and sends them to the $D$ distributed authorities. For example, some generic attributes of the user, such as, country, age, zipcode can be non-sensitive information that can be shared with multiple authorities. One of these organizations can serve as the central authority by verifying these attributes and relaying them to the rest, who verify only one sensitive attribute each of the user. This way, few potentially non-sensitive (or generic) attributes of the user are revealed to all the authorities, still preserving the privacy of the sensitive user attributes. This system is \emph{heterogeneous} both in terms of the verification responsibilities of the authorities, as well as the privacy requirements of the attributes. This natural heterogeneity in user attributes enables us to develop a hybrid scheme to improve the communication efficiency significantly, almost doubling the rate, and also decouple the number of attributes $N$ and the number of distributed servers $D$.

In an $(N,K)$ DAPAC system, we use the attribute vector ${\bm{v}^*}=(v_1^*,v_2^*,\ldots,v_N^*)$ to represent the user's $N$ independent attributes. Each attribute $v_n^*$ takes one of $K$ possible values from a corresponding finite alphabet set $\mathcal{V}_n$, with $|\mathcal{V}_n|=K$, $n\in [N]$, where $[N]:=\{1,\ldots,N\}$. The servers store a copy of the entire message set $\mathcal{W}$, where each message is identified by a unique attribute vector. The user commits $v_n^*$ to server $n$ and on verification, is given access to a subset of messages of $\mathcal{W}$. The user is allowed to correctly recover the designated message $W_{ {\bm{v}^*}}$, while acquiring no information on any other message, accessible or not. Based on the user's query, each server releases the requested combinations of accessible messages to the user, secured by a common randomness symbol shared among the servers. From the queries and answers, the user exactly recovers $W_{ {\bm{v}^*}}$, while gaining no information on $\mathcal{W}_{\bm{\bar{v}}^*}:=\mathcal{W}\backslash W_{ {\bm{v}^*}}$. The queries are designed such that server $n$ obtains no information on the user's attribute vector beyond $v^*_n$. The goal is to design a scheme that achieves this with a minimum cost of downloading answers per symbol of $W_{\bm{v}^*}$. Equivalently, the rate of an $(N,K)$ DAPAC system, defined as the ratio of message length to the download cost, should be maximized. Reference \cite{aliDAPAC} shows that the rate $\frac{1}{2K}$ is achievable. We illustrate this with an example.

\begin{example} \label{example1}
    Consider a $(3,2)$ DAPAC system with $N=3$ attributes ${\bm{v}}=(v_1,v_2,v_3)$ with $v_1 \in \mathcal{V}_1=\{a,b\}$, $v_2 \in \mathcal{V}_2=\{1,2\}$ and $v_3 \in \mathcal{V}_3=\{x,y\}$ and each message having $L=3$ symbols from $\mathbb{F}_q$. Let ${\bm{v}^*}=(v_1^*,v_2^*,v_3^*)=(a,2,y)$ be the particular attribute of the user. Thus, the designated message for the user is $W_{a2y}$. The user sends $v_1^*=a$ to server 1, $v_2^*=2$ to server 2, and $v_3^*=y$ to server 3. Once verified, the messages eligible for the user to access at each server are,
    \begin{align}
      \text{Server 1:} & \qquad \{W_{a1x}, W_{a1y}, W_{a2x}, W_{a2y}\}, \nonumber \\
      \text{Server 2:} & \qquad \{W_{a2x}, W_{a2y}, W_{b2x}, W_{b2y}\}, \nonumber \\
      \text{Server 3:} & \qquad \{W_{a1y}, W_{a2y}, W_{b1y}, W_{b2y}\}.
    \end{align}
    
    The user generates $12$ random vectors $h_{nm} ,\ n\in[3], \ m\in [4]$ uniformly from $\mathbb{F}_q^2$, $9$ of which are independent, and  $3$ of which satisfy, $h_{21}=h_{12}+e_2$, $h_{31} = h_{14}+e_2$ and $h_{34} = h_{24}+ e_1$, with $e_m$ denoting the $2\times 1$ column vector with zeros in all but the $m$th row. As a query to server $n$, user sends $h_{nm}, \ m\in [4]$ and the permuted indices of symbols to be linearly combined with $h_{nm}$. All operations are performed in $\mathbb{F}_q$. The answers downloaded from  each server is shown in Table~\ref{tabex12}. Here, $s_j \in \mathbb{F}_q, \ j\in[9]$ are independent parts of the shared common randomness.
    
    \begin{table}[htbp]
    \centering
    \begin{tabular}{|c|c|}
    \hline
    Server & Answer\\
    \hline
    \multirow{4}{*}{DB 1} & $h_{11} ^T [w_{a1x}(1) ; w_{a1y}(1)] +s_1$ \\ 
    & $h_{12} ^T [w_{a2x}(1) ; w_{a2y}(1)]+s_2$ \\ 
    & $h_{13} ^T [w_{a1x}(2) ; w_{a2x}(2)] +s_3$ \\ 
    & $h_{14} ^T [w_{a1y}(2) ; w_{a2y}(2)] +s_4$ \\
    \hline
    
    \multirow{4}{*}{DB 2} & $(h_{12} +e_2) ^T [w_{a2x}(1) ; w_{a2y}(1)] +s_2$ \\ 
    & $h_{22} ^T[w_{a2x}(3) ; w_{b2x}(1)]+s_5$ \\ 
    & $h_{23} ^T [w_{b2x}(2) ; w_{b2y}(1)] +s_6$ \\ 
    & $h_{24} ^T [w_{a2y}(3) ; w_{b2y}(2)] +s_7$ \\
    \hline
    
    \multirow{4}{*}{DB 3} & $(h_{14} + e_2)^T [w_{a1y}(2) ; w_{a2y}(2)] +s_4$ \\ 
    & $h_{32} ^T [w_{a1y}(1) ; w_{b1y}(1)]+s_8$ \\ 
    & $h_{33} ^T [w_{b1y}(2) ; w_{b2y}(2)] +s_9$ \\ 
    & $(h_{24} +e_1) ^T [w_{a2y}(3) ; w_{b2y}(2)] +s_7$ \\
       \hline
    \end{tabular}
    \vspace*{0.2cm}
    \caption{Answers downloaded from servers.}
    \label{tabex12}
    \vspace*{-0.4cm}
    \end{table}
    
    By subtracting the second answer of server 1 from the first answer of server 2, the user obtains the symbol $w_{a2y}(1)$ of $W_{a2y}$. The two remaining symbols are similarly obtained by pairing answers of servers $1$, $3$ and servers $2$, $3$. Thus, to recover $3$ message symbols, the user downloads $12$ symbols, yielding the $(3,2)$ DAPAC rate of $\frac{1}{4}$ which is $\frac{1}{2K}$, with $K=2$.  
\end{example}

In general, in an $(N,K)$ DAPAC system, the achievable scheme splits each message into $\binom{N}{2}$ equal chunks (sub-packets). The user sends queries to each server requesting linear combinations of sub-packets accessible from that server, such that, the privacy of the attributes other than the one verified is maintained. The user makes $K(N-1)$ downloads from each server, which are designed to preserve the secrecy of all but the designated message. Using the $KN(N-1)$ downloads, the user recovers $\binom{N}{2}$ symbols of the designated message. The achievable rate is therefore $\frac{N(N-1)/2}{KN(N-1)}=\frac{1}{2K}$.

Our main contribution is to leverage the natural heterogeneity in the user attributes and prove that, by employing a central server to verify and relay the generic (or non-sensitive) attributes, the achievable rate can be almost doubled from $\frac{1}{2K}$ to $\frac{1}{K+1}$. The scheme that achieves this rate results in asymmetric download costs from the central and the dedicated servers. This can be circumvented, if desired, by time-sharing between our $(N,D,K)$ HetDAPAC scheme and \cite{aliDAPAC}'s $(D,K)$ DAPAC scheme, with a higher total download cost. Further, by identifying a special property of the $(N,3,K)$ HetDAPAC system, we propose a new achievable scheme for this class.

\section{System Model}
\begin{figure}[t]
    \centering
    \includegraphics[width=0.7\linewidth]{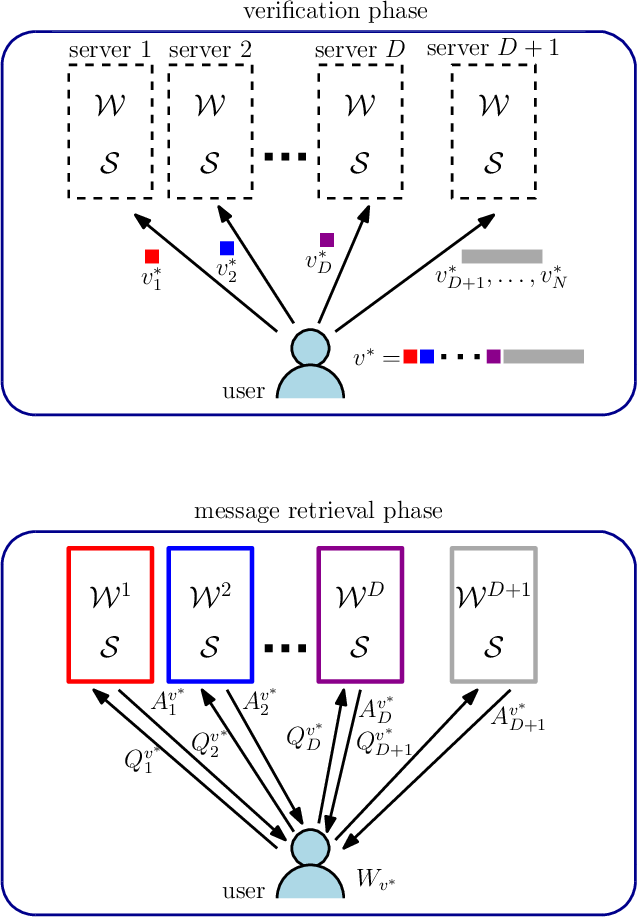}
    \caption{Formation of an $(N,D,K)$ HetDAPAC system.}
    \label{model}
    \vspace*{-0.4cm}
\end{figure}

We consider the setup with $(D+1)$ servers as shown in Fig.~\ref{model}. Similar to the $(N,K)$ DAPAC system, the user is identified by $N$ attributes and wishes to access its designated record stored in multiple servers. Different from \cite{aliDAPAC}, while we allow a subset of $D$ out of $N$ attributes to be verified by dedicated distributed servers (one attribute verified per server), the remaining $(N-D)$ attributes are verified by the $(D+1)$st server, which is central. The $(N-D)$ attributes, after verification, are conveyed by the central server to the $D$ dedicated servers. Thus, in our $(N,D,K)$ HetDAPAC system, $(N-D)$ attributes are publicly known to all $(D+1)$ servers in the system, while each dedicated server learns and verifies only one additional attribute. If $N-D=0$, by omitting the central server, our $(N,D,K)$ HetDAPAC system reduces to the $(N, K)$ DAPAC system. 

The servers store the message set $\mathcal{W}$ of $K^N$ messages in a replicated manner, each message $W_{\bm{v}}$ corresponding to a unique attribute vector $\bm{v}$ along with a set of common randomness $\mathcal{S}$, independent of ${\bm{v}}^*$, $\mathcal{W}$ and is unknown to the user. Each $W_{\bm{v}}$ is an independent message of $L$ symbols from a finite field $\mathbb{F}_q$. The symbols in $\mathcal{S}$ are uniformly and independently picked from $\mathbb{F}_q$. Under this setup, the user wishes to download its designated message $W_{\bm{v}^*}$, corresponding to its attribute vector $\bm{v}^*$. This is performed in two phases:

\subsubsection{Verification Phase} The user sends its attributes $v^*_{n},n\in [D]$ to server $n$ and the remaining attributes, $v^*_{D+1}, \ldots, v^*_N$, to server $(D+1)$ for verification. Server $(D+1)$ verifies the $(N-D)$ attributes and declares them to the $D$ servers, while remaining oblivious to the attribute values $v^*_1, \ldots, v^*_D$. 

\subsubsection{Message Retrieval Phase} Note that, in server $(D+1)$, there are $K^D$ accessible messages, corresponding to attribute vectors whose attribute values $v_{D+1},\ldots,v_N$ are fixed to $v^*_{D+1},\ldots,v^*_N$; and in server $n\in[D]$, there are $K^{D-1}$ messages corresponding to attribute vectors for which $v_n = v^*_n$ and $v_{D+1},\ldots,v_N=v^*_{D+1},\ldots,v^*_N$. We call the message repository in server $n$, as database $n$, and represent it by $\mathcal{W}^n$. 

The access control imposes a non-colluding \emph{non-replicated} database structure, with the following properties: i) only the designated message $W_{\bm{v}^*}$ is replicated in all databases, ii) any pair of dedicated databases share $K^{D-2}$ common messages, iii) the content of any database-$n$ is a subset of the content of database-$(D+1)$, for $n\in [D]$. While properties i) and ii) are applicable to an $(N,K)$ DAPAC system with $D=N$, iii) occurs additionally in our proposed $(N,D,K)$ framework.

Given the database structure, the user sends queries $Q_n^{\bm{v}^*}$ to server $n\in [D+1]$. The queries are generated without any knowledge of the contents of the databases, thus,
\begin{align}
     I(Q^{\bm{v}^*}_1,\ldots,Q^{\bm{v}^*}_{D+1}; \mathcal{W})=0.
\end{align}
To each dedicated server $n\in [D]$, no information on the $(D-1)$ remaining attributes of the user should be leaked from the queries. If $\bm{v}^*$ denotes the random attribute vector, then,
\begin{align}\label{user_priv1}
I(Q^{\bm{v}^*}_n; v^{*}_{[1:D]}\backslash v^{*}_n|\mathcal{S}, v^{*}_{[D+1:N]},v^{*}_n) = 0.
\end{align}
Similarly, to the $(D+1)$st server, user queries should reveal no information on any other attribute,
\begin{align}\label{user_priv2}
  I(Q^{v^*}_{D+1}; v^{*}_{[1:D]}|\mathcal{S},v^{*}_{[D+1:N]}) =0.  
\end{align}
Here, \eqref{user_priv1} and \eqref{user_priv2} are the \emph{attribute privacy} constraints, which allow the leakage of the last $(N-D)$ attributes to all the servers. In response to the queries, each server $n\in [D+1]$ sends answer $A^{\bm{v}^*}_n$ which is a function of $Q^{\bm{v}^*}_n$, $\mathcal{W}^n$ and $\mathcal{S}$, 
\begin{align}
    H(A_n^{\bm{v}^*}|Q^{\bm{v}^*}_n, \mathcal{W}^n, \mathcal{S}) = 0.
\end{align} 
The queries sent and the answers generated should be such that the correctness, attribute privacy and database secrecy constraints are satisfied. For \emph{correctness}, the designated message $W_{\bm{v}}^{*}$ should be recoverable using all the queries and answers,
\begin{align}
    H(W_{\bm{v}^{*}}|Q^{\bm{v}^*}_{[1:D+1]}, A^{\bm{v}^*}_{[1:D+1]})=0.
\end{align}
To ensure \emph{database secrecy}, the user should not learn anything about the remaining messages in any database beyond the message it is supposed to learn,
\begin{align}
    I(\mathcal{W}_{\bm{\bar{v}}^*}; Q^{\bm{v}^*}_{[1:D+1]},A^{\bm{v}^*}_{[1:D+1]}) =0.
\end{align}
The \emph{rate} of an $(N,D,K)$ HetDAPAC system is the ratio of the message length $L$ to the total download cost from all servers. Further, since our scheme results in asymmetric download costs between the first $D$ servers and the $(D+1)$st server, we quantify this asymmetry by a metric called \emph{load ratio}, denoted by $\ell$, which is defined as the ratio of the download costs from any server in $[D]$ and the server $(D+1)$.  

\section{Main Results}

\begin{theorem}\label{thm1}
    For an $(N,D,K)$ HetDAPAC system, the following rate $R$ is achievable,
    \begin{align}
        R=\frac{1}{K+1},
    \end{align}
    at a load ratio of $\frac{1}{KD}$ with a minimum required common randomness size $H(\mathcal{S})\geq KL$.
\end{theorem}

The proof of Theorem~\ref{thm1} and the subsequent remarks follow from the achievable scheme in Section~\ref{ach1}. 

\begin{remark}\label{rmk1}
    Similar to the scheme in \cite{aliDAPAC}, the rate of our scheme is independent of the number of attributes $N$, and the number of databases $(D+1)$, and decreases with the attribute alphabet size $K$.
\end{remark}

\begin{remark}
    If server $(D+1)$ participates only in the verification phase, and not in the message retrieval phase, then the scheme in \cite{aliDAPAC} is a valid achievable scheme for our problem with $N=D$. The resulting rate is $\frac{1}{2K}$, at a load ratio $\infty$, with a minimum common randomness requirement of $K^2L$.    
\end{remark}

\begin{remark}\label{rmk3}
    By time-sharing between our achievable scheme and that of a $(D,K)$ DAPAC system, any rate value in $\big(\frac{1}{2K}, \frac{1}{K+1}\big)$ at a load ratio in $\big(\frac{1}{KD},\infty\big)$ is achievable. 
\end{remark}

\begin{theorem}\label{thm2}
    For the class of $(N,3,K)$ HetDAPAC systems, the rate $\frac{2}{3K}$ is achievable, at a load ratio of $\frac{2}{3}$ with a minimum common randomness size of $\frac{K^2L}{2}$ symbols.
\end{theorem}

The proof of Theorem~\ref{thm2} follows from the new achievable scheme in Section~\ref{achnew}. 

\section{Achievable Scheme}\label{ach1}

\subsection{Motivating Example}
We illustrate the achievable scheme with a $(3,2,2)$ HetDAPAC system, with the same attributes as in Example~\ref{example1}.

\begin{example}
    Let each message consist of $L=2$ symbols in $\mathbb{F}_q$ and $\bm{v}^*=(a,2,y)$. After the verification phase, the messages accessible to the user from the databases are,
    \begin{align}
    \text{DB 1} &= \{W_{a1y}, W_{a2y}\}, \nonumber \\
    \text{DB 2} &= \{W_{a2y}, W_{b2y}\}, \nonumber \\
    \text{DB 3} &=\{W_{a1y}, W_{a2y}, W_{b1y}, W_{b2y}\}.
    \end{align}
    
    The user chooses a private permutation $\pi_{\bm{v}}$ to rearrange the symbols of  $W_{\bm{v}}$, with $[w_v(1) \, w_v(2)]$ denoting the permuted symbols. The user generates $4$ independent random vectors $h_{11}, h_{12}, h_{21}, h_{22}$ uniformly from $\mathbb{F}_q^{2\times 1}$. From DB 3, the user requests four message groups $w_{nk}$ where $ n, k \in [2]$ indicating the symbols to be linearly combined with $h_{nk}$. From DB 1 and DB 2, only $w_1:=w_{11}$ and $w_2:=w_{22}$, i.e., the message groups containing a symbol of $W_{a2y}$ are requested. The answers sent in response by each server is given in Table~\ref{tabex22}. The servers secure each answer with an independent part $s_{nk}$ of $\mathcal{S}$ shared among them to respect database secrecy.
    
    \begin{table}[htbp]
    \centering
    \begin{tabular}{|c|c|c|}    
    \hline
    Server & Message group & Answer\\
    \hline
    DB 1 & $ w_{1} = [w_{a1y}(1) ; w_{a2y}(1)]$ & $( {h_{11}} + {e_2})^T w_{1}  + s_{11} $\\
    \hline
    DB 2 & $ w_2 = [w_{a2y}(2) ; w_{b2y}(2)]$ & $( {h_{22}}+ {e_1}) ^T  w_{2}+ s_{22} $ \\
    \hline
     & $w_{11}=[w_{a1y}(1) ; w_{a2y}(1)]$ &  $ {h_{11}}^Tw_{11}  + s_{11} $\\
     DB 3 & $w_{12}=[w_{b1y}(1) ; w_{b2y}(1)]$ & ${h_{12}}^T w_{12} + s_{12} $\\
    & $ w_{21}=[w_{a1y}(2) ; w_{b1y}(2)]$ &  $ {h_{21}}^T w_{21}  + s_{21} $\\
    & $ w_{22}=[w_{a2y}(2) ; w_{b2y}(2)]$ &  $ {h_{22}}^T w_{22}  + s_{22} $\\
    \hline
    \end{tabular}
    \vspace*{0.2cm}
    \caption{Queries and answers for $ {\bm{v}}^*=(a, 2, y)$.}
    \label{tabex22}
    \vspace*{-0.4cm}
    \end{table}
    
    The user downloads $1$ symbol each from servers 1 and 2, and $4$ symbols from server 3, i.e., $6$ symbols total to recover $2$ symbols of $W_{a2y}$, yielding a rate $\frac{1}{3}$, i.e., $\frac{1}{K+1}$, load ratio $\frac{1}{4}$ and using $4$ symbols of common randomness. Since servers 1 and 2 are each requested for a single symbol of the user-accessible messages, while server 3 is requested for two symbols of its accessible database, attribute privacy is preserved. 
\end{example}

\subsection{Proof of Theorem 1}
Now, we describe the achievable scheme resulting in the rate $\frac{1}{K+1}$. Each message in $\mathcal{W}$ is split into $D$ equal sub-packets of $L/D$ symbols each. Thus, $W_{\bm{v}}= [W_{\bm{v}}(1) \ W_{\bm{v}}(2) \ \ldots \ W_{\bm{v}}(D)]$ where each sub-packet $W_{\bm{v}}(i)$ is a row vector of $L/D$ symbols from $\mathbb{F}_q$. For each accessible message corresponding to ${\bm{v}}$, the user privately chooses an independent random permutation $\pi_{{\bm{v}}}$ from  the set of all permutations $\Pi$, $|\Pi|=D!$ on $[D]$ to rearrange the sub-packets. Let $w_{\bm{v}}(i)= W_{\bm{v}}(\pi_{\bm{v}}(i))$ and $[w_{\bm{v}}(1) \ w_{\bm{v}}(2) \ \ldots \ w_{\bm{v}}(D)]$ denote the permuted sub-packets. Similarly, $\mathcal{S}$ is split into chunks of $L/D$ symbols from $\mathbb{F}_q$.

\subsubsection{Query Construction} We represent each attribute alphabet set $\mathcal{V}_n$ as an ordered list, where $\mathcal{V}_n(k)$ is the $k$th value in the list. This order is known to the user and the servers. We begin with the queries sent to server $(D+1)$. For each $n\in[D]$, $k\in [K]$, consider the collection of sets,  
\begin{align}
    \mathcal{U}(n,k)=\{W_{\bm{v}}\mid v_n=\mathcal{V}_n(k), \ v_{[D+1:N]}=v^*_{[D+1:N]}\}.
\end{align}

Clearly, $|\mathcal{U}(n,k)|=K^{D-1}$. For each message in $\mathcal{U}(n,k)$, the user chooses a new sub-packet index and vertically concatenates these sub-packets to obtain $w_{nk}$. Each $w_{nk}$, composed of $K^{D-1}$ sub-packets from accessible messages, is a $K^{D-1}\times (L/D)$ matrix in $\mathbb{F}_q$. The user generates $KD$ random vectors $h_{nk}, n\in [D]$, $k\in [K]$ of dimension $K^{D-1}\times 1$, whose elements are uniformly and independently picked from $\mathbb{F}_q$.

Note that, $\mathcal{W}^n \in \mathcal{U}(n,k)$ and let $k^*$ be such that for $n\in[D]$, $\mathcal{W}^n = \mathcal{U}(n,k^*)$, where for ${\bm{v}}=(v_1,\ldots, v_N)$,
\begin{align}
    \mathcal{W}^n &=\{W_{\bm{v}}, {\bm{v}} \mid v_n=v_n^*, \ v_{[D+1:N]}=v^*_{[D+1:N]}\}.
\end{align}
For the dedicated servers $n\in[D]$, define
\begin{align}
   w_{n}=w_{nk^*}, \qquad h_n=h_{nk^*} + e_{k^*}.
\end{align}
Thus, the query tuple $Q_n^{{\bm{v}}^*}$ for server $n$ is,
\begin{align}
   \!Q_n^{{\bm{v}}^*} \!=\! 
    \begin{cases}
        (h_n, w_n), & \!\!n\in [D]\\
        \{(h_{mk},w_{mk}), m\in [D], k\in [K]\}, & \!\!n=D+1.
    \end{cases}\!\!
\end{align}

\subsubsection{Answer Generation}  The servers collectively assign one chunk of $L/D$ symbols of $\mathcal{S}$ to each $\mathcal{U}(n,k)$, $n\in[D]$, $k\in [K]$ and label this as $s_{nk}$. To generate answers, the servers multiply (in $\mathbb{F}_q$) the entries in the query tuple  and add a chunk of $\mathcal{S}$ to it, as given below,
\begin{align}
    \!\!\!\!A_n^{\bm{v}^*}\!=\! 
    \begin{cases}
        h_n^T w_n +s_{nk^*}, & \!\!n\in [D]\\
        \{h_{mk}^T w_{mk} + s_{mk}, m\!\in\![D], k\!\in\![K]\}, & \!\!n=D\!+\!1.
    \end{cases}\!\!\!
\end{align}
To obtain $w_{\bm{v}^*}(n)$, the user subtracts the answer corresponding to the query $(h_{nk^*}, w_{nk^*})$ to server $(D+1)$ from the answer of the dedicated server $n$. This way, the $D$ sub-packets of $W_{\bm{v}^*}$ are recovered and rearranged with $\pi_{\bm{v}^*}^{-1}$. The resulting rate is,
\begin{align}
    R&=\frac{D}{KD+D}=\frac{1}{K+1}.
\end{align}
Since the user downloads one linear combination of sub-packets from each dedicated server, and $KD$ linear combinations from server $(D+1)$, the resulting load ratio $\ell$ is $\frac{1}{KD}$.

Each dedicated server uses the coefficients $h_n$ to linearly combine the sub-packets in $w_n$. Thus, the user downloads one sub-packet each of all the messages it is accessible to, irrespective of the values of the remaining attributes. From the central server, the user downloads $KD$ linear combinations of requested sub-packets, independent of $v^*_{[1:D]}$. Thus, attribute privacy is preserved. Since all the remaining responses from the central server are protected by symbols of $\mathcal{S}$, data secrecy is also preserved. For each $\mathcal{U}(n,k)$, $\frac{L}{D}$ distinct symbols of $\mathcal{S}$ are used, thus $H(\mathcal{S})\geq KL$.

\subsection{Proof of Remark~\ref{rmk3}}
 For a given $\lambda\in(0,1)$, we apply the scheme in \cite{aliDAPAC} to the first $\lambda L$ symbols and the scheme in Section~\ref{ach1} to the remaining $(1-\lambda)L$ symbols of each message to obtain a valid $(N,D,K)$ HetDAPAC scheme. Then, the download cost from each dedicated server is $\lambda \frac{2LK(D-1)}{D(D-1)}+ (1-\lambda)\frac{L}{D}$ and that from server $(D+1)$ is $(1-\lambda)KD\frac{L}{D}$. Thus, the achievable rate as a function of $\lambda$ is,
\begin{align}
    R(\lambda)= \frac{1}{K(1+\lambda)+(1-\lambda)}. \label{rate_lambda}
\end{align}
The load ratio $\ell(\lambda)$ as a function of $\lambda$ is,
\begin{align}
    \ell(\lambda) = \frac{2K\lambda + 1 - \lambda}{KD(1-\lambda)}=\frac{1}{KD} + \frac{2\lambda}{D(1-\lambda)}. \label{load_ratio_lambda}
\end{align}
Formally, the trade-off between rate and load ratio is,
\begin{align}
    R(\lambda) = \frac{1}{K+\frac{\ell(\lambda)K^2D+K}{\ell(\lambda)KD+2K-1}}.
\end{align}

\begin{remark}
In the next section, we propose a scheme for the class of $(N,3,K)$ HetDAPAC systems which, at a load ratio of $\frac{2}{3}$ attains the rate $\frac{2}{3K}$. On the other hand, the time-shared achievable rate under the same load ratio,  with $\lambda = \frac{2K-1}{4K-1}$ is $R(\lambda)= \frac{4K-1}{6K^2}$, from \eqref{rate_lambda}. The new scheme strictly outperforms the time-sharing scheme in terms of rate, since $\frac{2}{3K}>\frac{4K-1}{6K^2}$.
\end{remark}

\section{New Scheme for $(N,3,K)$ HetDAPAC System}\label{achnew}
 For $D>2$, any pair of dedicated servers have $K^{D-2}>1$ common messages. This observation inspired the scheme in \cite{aliDAPAC}. In our setup, we can leverage the message overlap between dedicated servers, simultaneously with the central server if $D=3$ and recover $2$ sub-packets from every pair of dedicated servers, and the central one (server 4), motivating our scheme for $(N,3,K)$ HetDAPAC system. We demonstrate this with a $(4,3,2)$ HetDAPAC system. 

\begin{example} 
    Let $\bm{v}^* = (a,2,u, y)$ with $\mathcal{V}_3 = \{u,v\}$ and $\mathcal{V}_1=\{a,b\}$, $\mathcal{V}_2=\{1, 2\}$ and $\mathcal{V}_4=\{x, y\}$ as before. The messages, each of length $L$, are split into $3(3-1) = 6$ sub-packets and permuted with independent user-private permutations. Once $v^*_{[D+1:N]}$ (here $v^*_4$) is verified and learnt, all the databases collectively assign $L/6$ symbols of the common randomness $\mathcal{S}$ to each message subset sharing $2$ common attributes in $v_{[1:3]}$ with $v_{[4:N]} = v^*_{[4:N]}$. This requires $\binom{3}{2}K^2\times(L/6)=\frac{K^2L}{2}$ symbols of $\mathcal{S}$. After $v^*_n, n\in[3]$ is verified, DB $n$ has $2K$ such message subsets, accessible to the user, each with $K$ messages.
    
    Next, we assign sub-packet indices to these message subsets in order from DB 1-3. We organize the subsets in DB $n\in[3]$ into $2$ ordered sets $\mathcal{G}_{n,1}$ and $\mathcal{G}_{n,2}$ of $K$ message subsets, where each $\mathcal{G}_{n,i}$, $i=1,2$ covers the subset of $\mathcal{W}^4$ with the $n$th attribute fixed to $v^*_n$. Here, $\mathcal{G}_{1,1} = \{(W_{a1uy}, W_{a1vy}), (W_{a2uy}, W_{a2vy})\}$ and $\mathcal{G}_{1,2} = \{(W_{a1uy}, W_{a2uy}), (W_{a1vy}, W_{a2vy})\}$. For DB 1, we pick a new (permuted) sub-packet index for every message in the subsets. Next, for DB 2, if a message subset already appeared in DB 1, we assign a new sub-packet index for $W_{\bm{v}^*}$ only, while those of the remaining messages in the subset are identical to those in DB 1. Finally, for DB 3, if a message subset already appeared in DB 1 or 2, we assign a new sub-packet index for $W_{\bm{v}^*}$, while those of the remaining messages in the subset are identical to those in DB 1 or 2. This way, all $6$ sub-packets of $W_{\bm{v^*}}$ have appeared in the message groups. For the remaining subsets, a new sub-packet index is assigned to each message. To complete the query tuple, the user generates $6K$ vectors $h_{nk}, n\in [3], k\in [2K]$ from $\mathbb{F}_q^{K\times 1}$. For a reappearing message group, the same $h_{nk}$ is reused. To form answers, the servers linearly combine the requested $w_{nk}$ with $h_{nk}$ and add to it an independent chunk of $\mathcal{S}$. The message groups requested to DBs 1-3 and the answers sent by them are given in Table~\ref{tabex3}. 
    
    \begin{table}[htbp]
    \caption{Queries and answers for $ {\bm{v}}^*=(a, 2,u, y)$ for servers 1,2,3.}
    \centering
    \begin{tabular}{|c|c|c|}
    \hline
    Server & Message group & Answer\\
    \hline
    \multirow{4}{*}{DB 1}& $w_{11}= [w_{a1uy}(1) ; w_{a1vy}(1)]$ &  ${h_{11}}^Tw_{11}  + s_{1} $\\
    & $w_{12}= [w_{a2uy}(1) ; w_{a2vy}(1)]$ &  ${{h_{12}}^T w_{12} + s_{2} }$\\
    & $ w_{13} = [w_{a1uy}(2) ; w_{a2uy}(2)]$ &  ${{h_{13}}^T w_{13} + s_{3} }$\\
    & $ w_{14}= [w_{a1vy}(2) ; w_{a2vy}(2)]$ &  ${h_{14}}^T w_{14} + s_{4} $\\
    \hline
    
    \multirow{4}{*}{DB 2} & $ w_{21}= [w_{a2uy}(3) ; w_{a2vy}(1)]$ &  $ {{h_{12}}^T w_{21}  + s_{2} }$\\
     & $ w_{22}= [w_{b2uy}(1) ; w_{b2vy}(1)]$ &  ${h_{22}}^T w_{22}+ s_{5} $\\
     & $ w_{23} = [w_{a2uy}(4) ; w_{b2uy}(2)]$ &  $ {{h_{23}}^T w_{23}  + s_{6} }$\\
     & $ w_{24} =  [w_{a2vy}(2) ; w_{b2vy}(2)]$ &  $ {h_{24}}^T w_{24} + s_{7} $\\
    \hline
    
     \multirow{4}{*}{DB 3} & $ w_{31} =  [w_{a1uy}(2) ; w_{a2uy}(5)]$ &  $ {{h_{13}}^T w_{31} + s_{3} }$\\
    & $ w_{32} = [w_{b1uy}(1) ; w_{b2uy}(3)]$ &  ${h_{32}}^T w_{32} + s_{9} $\\
    & $ w_{33}= [w_{a1uy}(3) ; w_{b1uy}(2)] $ &  $ {h_{33}}^T w_{33}  + s_{8} $\\
    
    & $ w_{34} =  [w_{a2uy}(6) ; w_{b2uy}(2)] $ &  $ { {h_{23}}^T w_{34} + s_{6} }$\\
    \hline
    \end{tabular}
    \label{tabex3}
    \end{table}

    \begin{table}[htbp]
    \caption{Queries and answers for $ {\bm{v}}^*=(a, 2,u, y)$ for server 4.}
    \centering
    \begin{tabular}{|c|c|}
    \hline
     Message group & Answer \\
     \hline
       $[w_{11}; w_{12}]$ &  {$h_{11}^T w_{11} + (h_{12}+e_1)^T w_{12} +(s_1 +s_2)$}\\
       $[w_{41}]$ & $ h_{41}^T w_{41}+ (s_5 + s_{10})$\\
       $[w_{23};w_{24}]$ & {$(h_{23}+e_1)^T w_{23} + h_{24}^T w_{24} +(s_6 +s_7)$}\\
       $[w_{42}]$ & $  h_{42}^T w_{42} + (s_8 +s_{11})$\\
       $[w_{31};w_{32}]$ &  {$(h_{13}+e_2)^T w_{31} + h_{32}^T w_{32} +(s_3 +s_9)$}\\
       $[w_{43}]$& $h_{43}^T w_{43} + (s_4 + s_{12})$\\
       \hline
    \end{tabular}
    \label{tabex3-1}
    \end{table}
    
    To DB 4, the requested message subsets are formed by first grouping the messages that share $1$ attribute, besides the known $(N-3)$ (here, $1$) attributes. If the message subset contains $W_{\bm{v}^*}$, with $v_n=v^*_n ,\ n\in [3]$ as the common attribute, we order each of them according to one of the two subsets between $\mathcal{G}_{n,1}$ and $\mathcal{G}_{n,2}$ such that the three chosen subsets are pairwise disjoint. In this example, $\mathcal{G}_{1,1}, \mathcal{G}_{2,2}, \mathcal{G}_{3,1}$ is a valid choice. Then, we assign the same sub-packet indices to the chosen message subsets as queried to DB $n$. Similarly, we use the same $h_{nk}$ as used for DBs 1-3 concatenated vertically resulting in $K^2\times 1$ vectors. If $W_{\bm{v}^*}$ does not appear in the message subset, we assign new sub-packet indices to each message in the subsets to form message groups as follows, 
    \begin{align}
    w_{41}& = [w_{b1uy}(3) ; w_{b1vy}(1); w_{b2uy}(4) ; w_{b2vy}(3)],\\
    w_{42}& = [w_{a1uy}(4) ; w_{b1uy}(4) ; w_{a1vy}(3); w_{b1vy}(2)],\\
    w_{43}& = [w_{a1vy}(4) ; w_{a2vy}(3); w_{b1vy}(3) ; w_{b2vy}(4)].
    \end{align}
    Corresponding to $w_{4i}, i\in [3]$ the user generates new $K^2\times 1$ vectors $h_{4i}$. The message groups requested from DB 4, along with its returned answers are tabulated in Table~\ref{tabex3-1}. Note that, the effective common randomness here is the sum of the $K$ chunks of $\mathcal{S}$ assigned to the message subsets in $[1:K],[K+1:2K], \ldots ,[K^2-K+1:K^2]$ entries of the message group. 
    
    Now, by summing the first two answers of DB 1 and subtracting it from the first answer of DB 4, the user decodes $w_{a2uy}(1)$. Again, by subtracting the second answer of DB 1 from the first answer of DB 2, the user obtains $h_{12}^T(w_{a2uy}(3)-w_{a2uy}(1))$ from which it recovers $w_{a2uy}(3)$. From each pair of DBs 1-3 along with DB 4, the user decodes all $6$ sub-packets.  In total, $18$ sub-packets are downloaded, yielding the rate $\frac{1}{3}$, while that by time-sharing is 
    $\frac{7}{24}$.

\end{example}


\bibliographystyle{unsrt}
\bibliography{reference}
\end{document}